\documentclass[12pt]{article}

\usepackage{sbc-template}

\usepackage{graphicx,url}
\usepackage[brazil]{babel}   
\usepackage[utf8]{inputenc}  

\title{Promoção da Equidade de Gênero na Programação Competitiva: Estratégias e Impactos das Ações Afirmativas nas Maratonas de Programação no Brasil}

\author {Crishna Irion\inst{1}, Camila da Cruz Santos\inst{1}, Luiz Cláudio Theodoro\inst{1}, \\Rafael Dias Araújo\inst{1}, João Henrique de Souza Pereira\inst{1}}
\address{Faculdade de Computação (FACOM) -- Universidade Federal de Uberlândia(UFU)\\Uberlândia -- MG -- Brasil
\email{ \{crishna, camilacruz, luiz.theodoro, rafael.araujo, joaohs\}@ufu.br }
}

\begin{document} 
\maketitle

\begin{abstract}
In the context of Computing, competitive programming is a relevant area that aims to have students, usually in teams, solve programming challenges, developing skills and competencies in the field. However, female participation remains significantly low and notably distant compared to male participation, even with proven intellectual equity between genders. This research aims to present strategies used to improve female participation in Programming Marathons in Brasil. The developed research is documentary, applied, and exploratory, with actions that generate results for female participation, with affirmative and inclusion actions, an important step towards gender equity in competitive programming.

\end{abstract}
     
\begin{resumo}

No contexto da Computação, a programação competitiva é uma área relevante que tem como objetivo que alunos resolvam, geralmente em equipes, desafios de programação desenvolvendo habilidades e competências na área. No entanto, a participação feminina permanece significativamente baixa e notavelmente distante em comparação com a masculina, mesmo comprovada a equidade intelectual entre gêneros. O objetivo desta pesquisa é apresentar estratégias utilizadas para melhorar a participação feminina nas Maratonas de Programação no Brasil. A pesquisa desenvolvida é  documental, aplicada e exploratória, com ações que geram resultados à participação feminina,  com ações afirmativas e de inclusão, um passo importante na direção da equidade de gênero na programação competitiva.

\end{resumo}

\section{Introdução}

A promoção da equidade de gênero em ambientes sub-representados por mulheres e predominante de homens é fundamental na atualidade. É reconhecido que as mulheres possuem capacidade cognitiva igual a dos homens\cite{difGenero}, entretanto, são menos vistas em profissões relacionadas à informática \cite{cooper2003}. Essa falta de representação é influenciada por diversos fatores de gênero, sociais e culturais \cite{Fisher2006}. No entanto, a discussão sobre igualdade de gênero tem ganhado destaque nos últimos anos, impulsionada por campanhas e movimentos de inclusão. Coletivos femininos e iniciativas acadêmicas estão se empenhando em atrair mais mulheres para a área de Tecnologia e Computação \cite{sbc22}. 

A educação emerge como um instrumento fundamental nesse processo, abrindo portas para conquistas em esferas sociais, econômicas e em campos específicos, como ciências exatas e tecnologia. O desenvolvimento de estratégias para abordar essas manifestações de desigualdade tem sido amplamente adotado no âmbito científico global, impulsionado principalmente pelos Objetivos de Desenvolvimento Sustentável (ODS) estabelecidos pela Organização das Nações Unidas (ONU) em 2015. O avanço da dimensão de gênero é fundamental para o avanço do conhecimento científico e para o bem-estar humano, incluindo o cumprimento das metas estipuladas pelos ODS. O Objetivo 5 dos ODS visa alcançar a igualdade de gênero e empoderar todas as mulheres e meninas, e inclui metas específicas relacionadas a esse propósito, visa``garantir a participação plena e efetiva das mulheres e a igualdade de oportunidades para a liderança em todos os níveis de tomada de decisão na vida política, econômica e pública'' \cite{ONU2030}.

A programação competitiva é uma área crescente que desafia programadores a resolver problemas computacionais complexos em um tempo limitado. Utiliza um conjunto de ferramentas lúdicas, aplicadas e eficazes para reunir estudantes e educadores de diversas áreas, submetendo-os a um envolvente sistema de competições \cite{wei5}. Essas competições não apenas abrangem o aprendizado técnico nas respectivas áreas do conhecimento, mas também cultivam uma série de habilidades complementares, incluindo o trabalho em equipe, a promoção de bons hábitos de estudo e a autonomia no processo de aprendizagem, bem como o desenvolvimento de capacidades para solução de problemas, controle e organização.

No contexto brasileiro, um leque diversificado de Olimpíadas Científicas abrange disciplinas como Química, Ciências, Matemática, Física, História, Astronomia, Informática, Biologia, entre outras, atendendo a diferentes níveis de ensino. Contudo, apesar da crescente importância desses campeonatos no cenário global e do crescente interesse dos estudantes, a participação feminina permanece notavelmente distante em comparação com a masculina \cite{wit1}. A falta de interesse  feminino  em  competições  de  programação acontece,  entre  outros  fatores, porque as competições não são pensadas, em sua essência, para participação  feminina. Elas podem se tornar mais inclusivas para as participantes do sexo feminino com a introdução de mudanças que as tornem mais sensíveis às diferenças baseadas no gênero \cite{Fisher2006}.

Uma grande disparidade nas participações pode ser observada ao analisar os níveis de ensino (fundamental, médio e superior) \cite{wit2}. Isso ressalta a necessidade de pesquisas que explorem esse fenômeno, buscando entender, avaliar e propor estratégias para equilibrar a participação de mulheres e homens nesse contexto. Algumas iniciativas visam quebrar esse paradigma como em ações desenvolvidas na Universidade de Brasília  \cite{wit3},  na Universidade Federal do Amazonas \cite{wit2} e, internacionalmente, com \cite{cooper2003} que discute o contexto feminino na Computação, assim como, \cite{Fisher2006} que pensam formas de mitigar a exclusão das meninas nas competições de programação no Canadá. As iniciativas que tem surgido até então, tem um teor regional ou fora do Brasil porém, no contexto nacional, esta pesquisa é pioneira.

O objetivo desta pesquisa é apresentar estratégias utilizadas para melhorar a participação feminina nas Maratonas de Programação. Para tanto foi desenvolvida uma pesquisa documental, seguida de análise do cenário feminino na programação competitiva, levantando dados sobre a participação em maratonas de programação e os resultados de um conjunto de ações afirmativas desenvolvidas na Maratona de Programação da Sociedade Brasileira de Computação (SBC) como estratégias para aumentar a participação feminina brasileira. 
 
Dessa forma, a importância desta pesquisa é trazer visibilidade a ações estratégicas para a equidade  de gênero. A compreensão do compromisso das universidades com a redução das desigualdades entre homens e mulheres deve vir acompanhada de uma série de medidas coordenadas em torno de um plano estratégico formal, de equidade de gênero, que permita o seu pleno desenvolvimento  em  todos  os  âmbitos  que  permeiam  a  profissão  acadêmica  e  a  organização  das universidades \cite{Clavero20}.

As contribuições do trabalho para a área de Computação e, especialmente para educação em Computação, são compreender cenário feminino na programação competitiva e os resultados de ações afirmativas e estratégias conduzidas para aumentar a participação feminina brasileira nas Maratonas de Programação realizadas pela SBC.

Este artigo está organizado da seguinte maneira: Introdução, seguida da Seção \ref{sec:firstpage},  uma visão geral dos trabalhos relacionados; na Seção \ref{sec:secondpage} é apresentado o método de pesquisa; na Seção \ref{sec:thirdpage} são analisados os dados coletados, resultados e a discussão em relação às Maratonas de Programação (SBC), referentes à primeira fase (regional) e na final nacional, mostrando o comparativo entre homens e mulheres. Em seguida na Seção \ref{sec:fourthpage} são apresentado resultados e impactos das ações afirmativas para participação feminina desenvolvidas nos últimos 2 anos; finalizando com as conclusões na Seção \ref{sec:conclusionpage}.

\section{Trabalhos Relacionados} \label{sec:firstpage}


\cite{AC_Scielo2023} repercutem em sua pesquisa a visibilidade do tema equidade  de gênero, mostrando que ainda é recente  no  Brasil. O trabalho mostra que o compromisso das universidades com a redução das desigualdades de gêneros precisa acompanhar um conjunto  de medidas associadas a um plano de equidade de gênero, que permita o desenvolvimento  em  todos  os  âmbitos  que  permeiam  a  profissão  acadêmica  e  a  organização das universidades.

 \cite{Patriarcado19} demonstra como os estereótipos de gênero são construídos socialmente e usados para justificar a dominação masculina. Sobre as mulheres na Computação, \cite{koch08} analisaram a atribuição do fracasso em Computação aos esteriótipos. Perceberam dois grandes aspectos que impactam na falta de sucesso: pouco conhecimento prático em informática e a condição de ameaça estereotipada, na qual a competência das mulheres com computadores é sempre questionada. Isto reflete nesta pesquisa, que mostra a baixa adesão feminina às competições de programação. O relatório anual da SBC (2022/2023) destaca que  está abaixo de 10\% o número de mulheres participantes das competições \cite{SBC2024}. 

\cite{Moro2022} defende que a integração de conceitos de diversidade, equidade e inclusão (DEI) sejam incorporados ao currículo da área de Computação. Em sua pesquisa, realizada com 118 docentes brasileiros, foi revelado que metade deles nunca abordaram DEI em suas aulas e demonstram resistência em discutir o tema. Isto reflete nesta pesquisa, que mostra a baixa adesão feminina às competições de programação. Iniciativas como a criação de estratégia para aumentar a participação feminina tem acontecido pontualmente em universidades, como na Universidade do Estado do Amazonas - UEA, que através do grupo de treinamento e pesquisa, tem obtidos resultados locais positivos \cite{wit1}. 

\cite{Fisher2006} argumentam que a estrutura atual das maratonas de programação pode ser desfavorável para mulheres, levando a uma menor participação e desempenho, afirmando que o gênero pode influenciar em como as pessoas se sairão em competições que, entre outras variáveis, a etnia, personalidade, experiência podem ser questões. Sugerem que a inclusão de mecanismos de \textit{Feedback}, a formulação de questões de forma a torná-las mais atraentes para mulheres, a criação de um ambiente mais privado e a implementação de sistemas de pontuação que valorizem a qualidade do código, além da velocidade, podem contribuir para um ambiente mais justo e inclusivo.

Os pesquisadores \cite{hello23} colocaram o tema DEI em pauta e promoveram a reflexão e conscientização sobre sua importância em seu trabalho. Propuseram articular cinco elementos como estratégia (tematizar, diversificar, espalhar, incrementar e explicitar) para incluir questões de gênero em Computação, principalmente o empoderamento feminino com propósito de promover a conscientização e empatia no
tema, e também a reflexão e a ação por meio da proposição de soluções computacionais. 

Algumas metodologias têm sido propostas, como \cite{wit2} e \cite{Bastos2017}, para aumentar o interesse, especialmente em alunas, nas maratonas de programação. Ambos os estudos têm como objetivo promover a inclusão de alunas nessas atividades e incentivar seu envolvimento através de diferentes abordagens. \cite{Bastos2017} desenvolveram uma abordagem mais abrangente, dividida em quatro partes: captação das alunas, treinamento das habilidades de programação, desenvolvimento contínuo dessas habilidades e, por fim, identificação de deficiências individuais de cada participante para personalizar o suporte e a orientação. Como resultados, perceberam que as mulheres em destaque no grupo de pesquisa permitem mudanças na visão da comunidade estudantil e acadêmica em geral, motivando  outros alunos e influenciando diretamente a determinação de que podem ter sucesso na área.

O trabalho de \cite{wit4} apresenta em sua análise a baixa participação feminina em competições de programação em nível local (região Norte do Brasil). O artigo destaca que a área da Computação, em geral, apresenta uma predominância masculina, tanto na escolha de cursos de graduação quanto na participação em eventos como maratonas de programação e \textit{hackathons}. O projeto propôs uma metodologia de envolvimento com palestras para despertar o interesse das alunas, seguidas de treinamento de programação, com destaque no envolvimento das alunas, tanto por meio de apresentações quanto de práticas concretas de codificação. Nos resultados das atividades realizadas por \cite{wit3} houve um aumento no número de meninas participantes das acadêmicas do projeto, em competições de programação em comparação a anos anteriores.

Isso sugere que as abordagens e metodologias implementadas foram eficazes em aumentar o interesse e a participação das alunas nessas competições. Porém, são iniciativas locais e pontuais. A Competição Feminina - Olimpíada Brasileira de Informática (OBI) - surgiu na pandemia e atende ao público feminino do ensino Médio e Fundamental. Vários projetos do Meninas Digitais atendem as universidades, mas não são específicos para programação competitiva. No Brasil, é a primeira vez que ações afirmativas são pensadas para atingir amplamente a programação competitiva em nível universitário, como as mostradas nesta pesquisa.

\section{Método de Pesquisa} \label{sec:secondpage}

Esta é uma pesquisa documental, aplicada e exploratória, em que a experiência dos autores com as ações executadas permitiu a análise qualitativa. Para \cite{cellard2012analise}, a análise documental permite examinar o desenvolvimento ou progresso de indivíduos, grupos, ideias, conhecimentos, comportamentos, mentalidades, práticas, e outros aspectos ao longo do tempo. O problema central da pesquisa consiste em analisar a participação das mulheres em competições de Programação Competitiva, avaliando a representatividade das mulheres nessas competições, propondo estratégias e analisando os resultados destas ações propostas.

\subsection{Revisão da literatura}

Para realização da pesquisa foi feita uma revisão bibliográfica de artigos com o tema  mulheres na Computação, mulheres nas olimpíadas de programação, mulheres e a programação competitiva. As pesquisas expuseram uma escassez de artigos voltados a participação feminina em competições de programação, principalmente um entendimento do cenário feminino nas Maratonas de Programação no contexto Nacional, o que reflete na importância de realizar este trabalho no Brasil.

Para isso, foram conduzidas buscas na BDTD (Biblioteca Digital Brasileira de Teses e Dissertações), bem como nas bases Springer, Scopus, IEEE, ACM e SBC SOL, com foco em trabalhos apresentados nos eventos WEI (\textit{Workshop} sobre Educação em Computação), WIT (\textit{Women in Technology}), CBIE (Congresso Brasileiro de Informática na Educação) e ANPED (Associação Nacional de Pós-Graduação e Pesquisa em Educação), ACM -\textit{Celebration of Women in Computing: womENcourage, Annual Conference on Research in Equity and Sustained Participation in Engineering, Computing, and Technology (RESPECT) e CMD-IT/ACM Richard Tapia Celebration of Diversity in Computing Conference (Tapia)}, trabalhos estes que se relacionam com a abordagem deste trabalho.

\subsection{Coleta e análise de dados}

Para atingir o objetivo proposto, foram coletados dados disponibilizados pelo comitê da Maratona de Programação (SBC) sobre as fases regionais e nacionais das maratonas nos últimos anos. Os dados públicos sobre estatísticas dos cursos superiores foram encontrados no site da SBC \cite{SBCestatisticas}, incluindo as estatísticas dos cursos de Computação de 2016 a 2021, os dados mais recentes publicados pela instituição.

Os dados das Maratonas de Programação do Brasil foram fornecidos pela Maratona SBC de Programação e são considerados fontes oficiais, conforme descrito pelo \textit{International Collegiate Programming Contest} (ICPC). Ao todo foram 29 maratonas de programação da SBC no Brasil e os dados consolidados estão disponíveis desde 2011 no site do ICPC global. Este trabalho abrange os anos de 2014 a 2023 e inclui informações sobre a primeira fase e a final nacional da Maratona de Programação no Brasil.

A análise estatística dos dados foi conduzida por meio de um algoritmo desenvolvido em \textit{Python}, permitindo a manipulação e avaliação das informações com precisão. Além disso, foram elaborados gráficos estatísticos utilizando software de planilhas para visualizar a análise estatística descritiva.

Um ponto importante a se compreender no cenário analisado é que o percentual de alunos de Computação que participam de Maratonas de Programação variam de 5\% a 8\%. Deste total de participantes, as meninas, que na Computação representam 13\% a 16\%, não representaram mais que 8\% de 2011 a 2021. 
Para que fique clara a estatística, a Tabela 1 apresenta o recorte amostral que a Maratona de Programação representa nos cursos de Tecnologia da Informação (TI). A percepção é ainda menor quando se observa a participação feminina nos cursos de tecnologia e seu recorte de participação nas Maratonas de Programação. 

\begin{table}[ht!]
\centering
\caption{Concluintes - cursos de TI x participantes das Maratonas}
\label{tab:exTable1}
\includegraphics[width=0.9\textwidth]{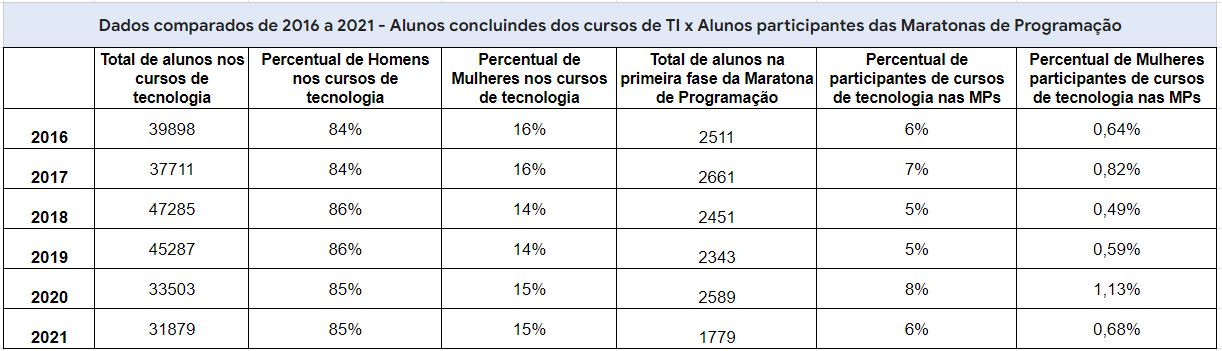}
\end{table}

\section{Análise dos resultados}\label {sec:thirdpage}

Esta pesquisa apresenta estratégias para melhorar a participação feminina nos cursos de Computação por meio das maratonas de programação. Para alcançar o objetivo foi necessário compreender o cenário feminino na programação competitiva e, posteriormente, analisar os resultados do conjunto de ações afirmativas desenvolvidas pelo comitê da Maratona de Programação (SBC), estratégias essas que buscam aumentar a participação feminina nas fases brasileiras e Latino-Americana da competição.

\begin{figure}[ht!]
\centering
\includegraphics[width=0.85\textwidth]{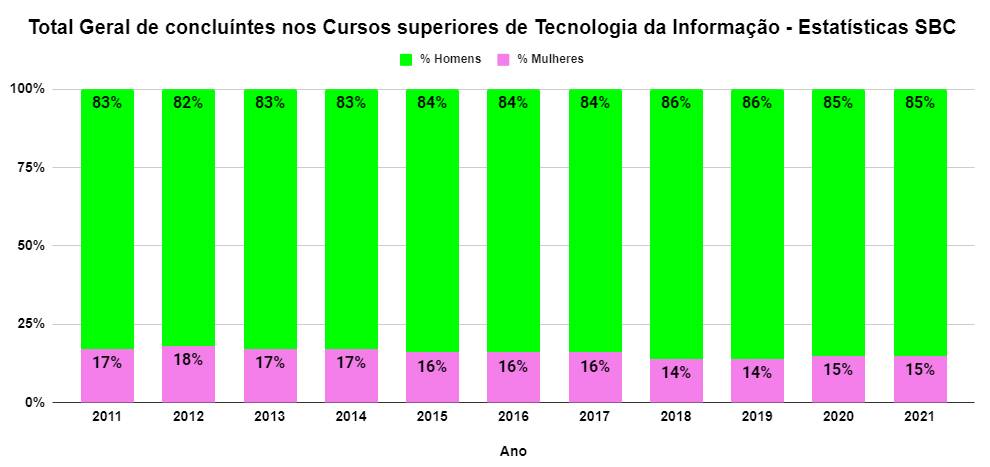}
\caption{Concluintes dos cursos de TI - 2011 a 2021}
\label{fig:exampleFig1}
\end{figure}

A primeira análise apresentada é o percentual de mulheres em todos os cursos da área de Computação e TI, pois todos cursos podem participar das Maratonas de Programação, não sendo exclusivo para cursos de Computação. As estatísticas anuais da SBC são apresentadas \cite{SBCestatisticas}, conforme 
a Figura~\ref{fig:exampleFig1}. Os dados são os disponíveis publicamente, não estando, até o momento, disponíveis os anos de 2022 e 2023.

Para \cite{Santos21}, em seu estudo sobre o cenário feminino nos cursos superiores da área de TI no Brasil, ao analisar os dados do INEP de 2014 a 2019, mostra que o número de discentes de tecnologia variou de 13,8\% a 15,2\% são mulheres cursando e formadas. A desigualdade de gênero é evidente tanto no número de pessoas cursando quanto no número de egressas. 

\begin{figure}[!ht]
\centering
\includegraphics[width=0.85\textwidth]{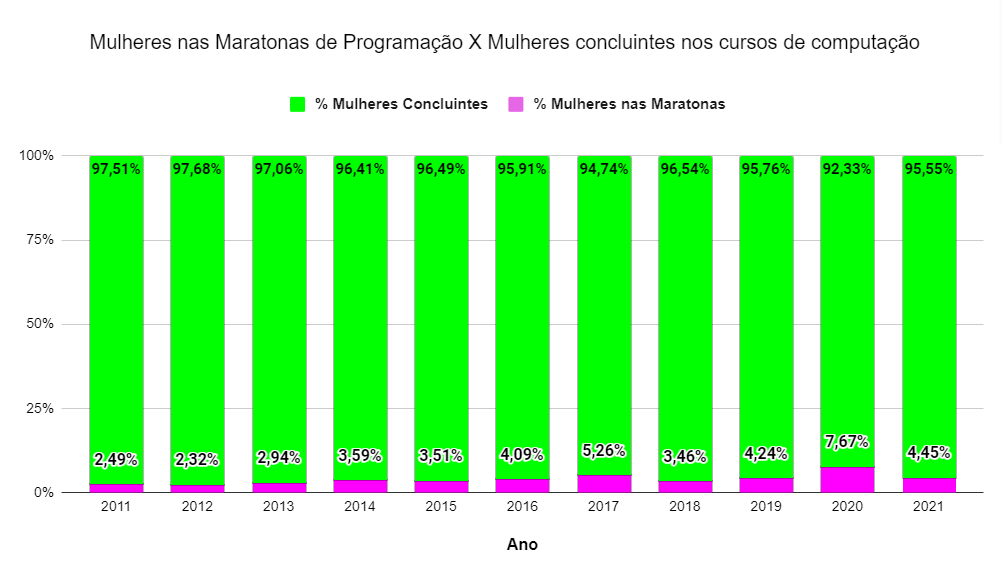}
\caption{Comparativo Mulheres Concluintes x Mulheres na primeira fase da Maratona de Programação}
\label{fig:exampleFig2}
\end{figure}

Já ao comparar as alunas concluintes nos cursos de Computação e TI, com as participantes das maratonas de programação, nota-se que é um pequeno percentual de alunas que está neste local de fala (Figura~\ref{fig:exampleFig2}). Quando o recorte é de participantes mulheres, este percentual é reduzido consideravelmente.

\begin{figure}[ht!]
\centering
\includegraphics[width=0.85\textwidth]{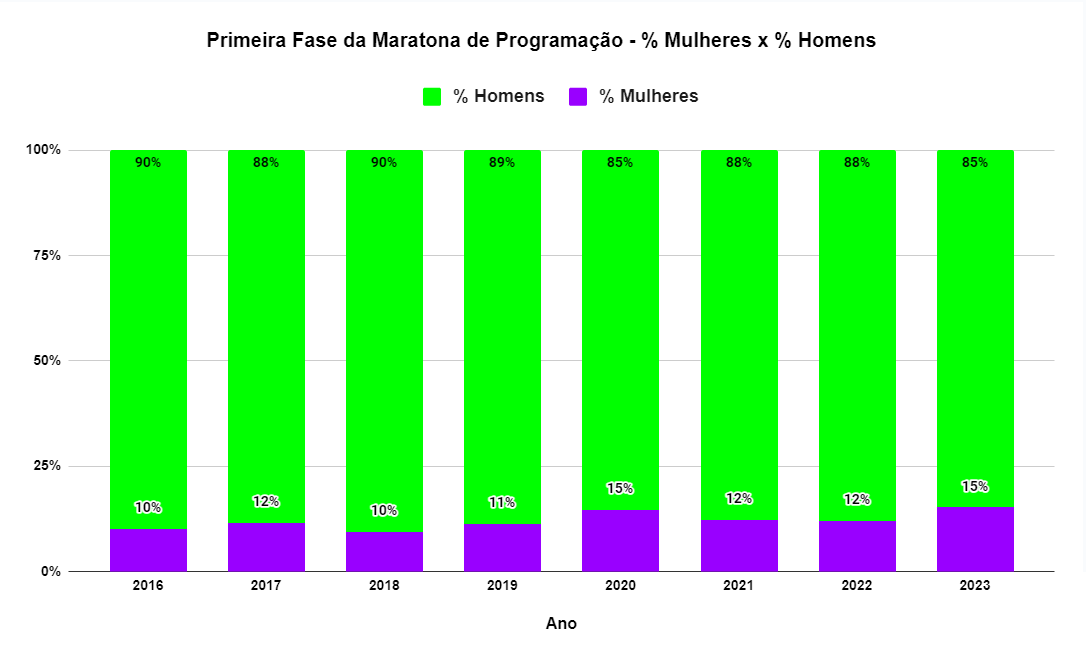}
\caption{Mulheres na Primeira Fase - 2016 a 2023}
\label{fig:exampleFig3}
\end{figure}

A participação das mulheres na Maratona de Programação, na primeira fase da maratona, que é de ampla concorrência, em que podem participar equipes de escolas do Brasil todo, sejam particulares ou públicas, é proporcional ao percentual de alunas nos cursos de Computação e TI. Em um comparativo dos últimos 5 anos percebe-se que somente nos anos de  2020 e 2023 houve um percentual maior que o de mulheres nos cursos de TI. A  Figura~\ref{fig:exampleFig3} mostra o percentual de participantes meninas por ano na primeira fase da Maratona de Programação SBC. Em um primeiro olhar, imagina-se que é um cenário positivo. Mas, 2020, um ano atípico, em que a prova da maratona foi realizada de forma remota, percebe-se um ponto interessante a ser analisado em trabalhos futuros sobre o impacto da pandemia no cenário feminino. Talvez a segurança de estar em casa, possa ter encorajado a participação. Já no ano de 2023, o aumento da participação começa a refletir as ações afirmativas que foram de fato implantadas e difundidas na Maratona.

\begin{figure}[ht!]
\centering
\includegraphics[width=0.85\textwidth]{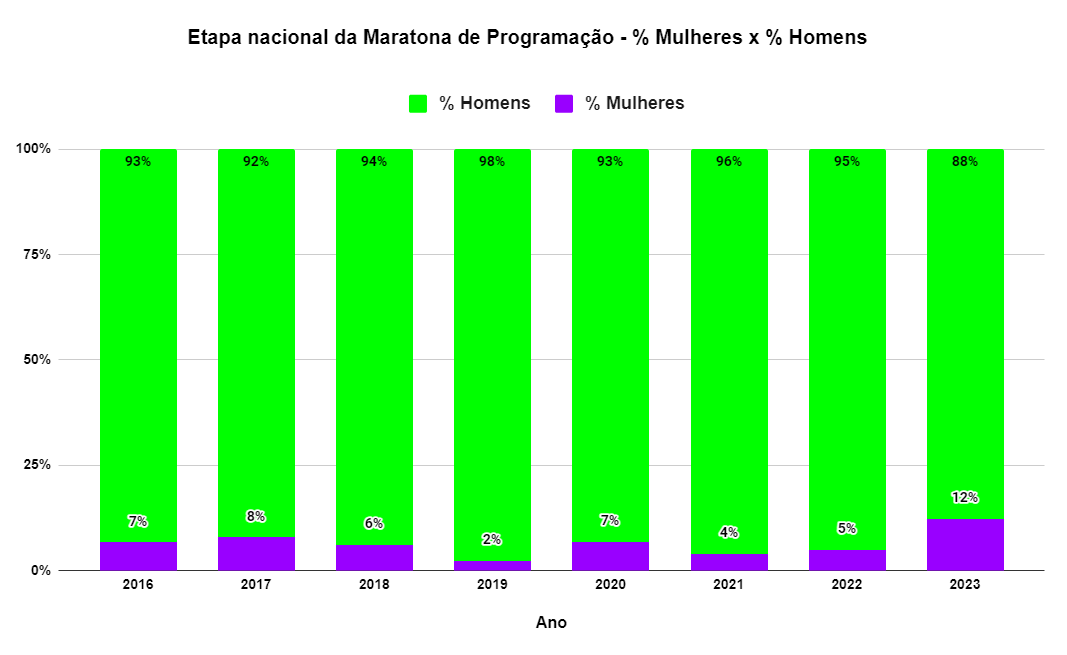}
\caption{Mulheres na Final Nacional - 2016 a 2023}
\label{fig:exampleFig4}
\end{figure}

Porém, quando se trata da final nacional da Maratona, a Figura~\ref{fig:exampleFig4} mostra um cenário bem mais significativo. O crescimento da participação feminina fica destacado na final nacional de 2023, em que as primeiras ações afirmativas foram implantadas. Foram oferecidas 14 vagas afirmativas e, com este incentivo, organicamente houve um aumento de outras 10 mulheres. Apesar de pequenos números absolutos, o crescimento percentual é relevante neste cenário, haja vista que em 2019 (antes da pandemia) apenas 2\% eram mulheres participantes.

\section{Discussões}\label{sec:fourthpage}

Iniciativas afirmativas são importantes em todas as áreas da sociedade a fim de encontrar a equidade. Projetos como ``Meninas Digitais'' têm desempenhado um papel fundamental na aproximação de meninas ao mundo da Computação \cite{SBC:23}. Além disso, coletivos em todo o Brasil têm desempenhado um papel crucial ao oferecer suporte e incentivo. Os resultados das análises revelam a necessidade de um esforço contínuo para melhorar a representatividade das mulheres na programação competitiva. Esse esforço deve ser multifacetado e envolver várias frentes de ação.

\begin{figure}[!ht]
\centering
\includegraphics[width=.85\textwidth]{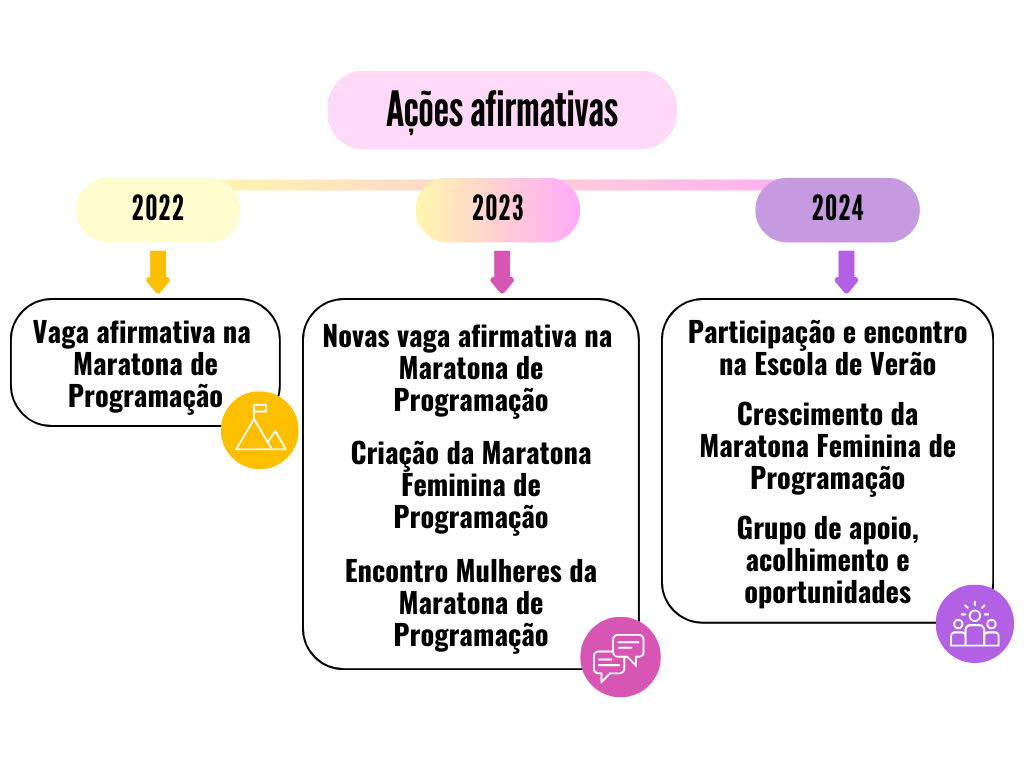}
\caption{Ações Afirmativas}
\label{fig:exampleFig5}
\end{figure}

Algumas ações afirmativas têm sido desenvolvidas para os resultados encontrados nesta pesquisa, conforme a Figura~\ref{fig:exampleFig5} . 

\noindent \textbf{\textit{Vaga afirmativa na Maratona de Programação:}} O comitê organizador da Maratona de Programação, na edição de 2021 da Final Nacional, adotou uma medida significativa, com a aprovação de uma vaga exclusiva para a melhor equipe de meninas no Brasil, para a participação na final de 2022. Como resultado, uma equipe de meninas alcançou o 36º lugar entre as 62 equipes participantes na final nacional e 618 equipes na primeira fase. 

Esta ação inspirou a criação da primeira Maratona Feminina de Programação (MFP - https://www.instagram.com/mfp.unicamp), realizada em 24 de junho de 2023, com o objetivo de encorajar a participação feminina (e não binária) nas competições de programação. A MFP oferece aulas online preparatórias, ministradas por mulheres renomadas na programação competitiva. Um total de 97 meninas de todo o Brasil participou da maratona proposta pelas meninas da equipe representativa. Já em 2024 a MFP tomou proporções ainda maiores, teve cerca de mil inscritas na primeira fase e 130 finalistas no evento organizado no mês de junho, com destaque para o grande engajamento da comunidade da Computação e, principalmente, da programação competitiva \cite{SBC2024}.

\noindent \textbf{\textit{Políticas de vagas extras:}} Como continuidade das ações afirmativas, o comitê da Maratona de Programação aprovou um conjunto de políticas para disponibilizar vagas extras para meninas e equipes com meninas em 2023. No relatório anual da SBC (2022/2023) a importância de aumentar a participação feminina em competições científicas é reconhecida e destaca a oferta de vagas para alunas nas escolas e, a partir de 2023, criou mais 5 vagas na final brasileira para times com participação feminina. Essas medidas visam elevar a participação feminina de menos de 10\% para cerca de 20\% dos competidores \cite{SBC2024}.

Os resultados se mostram promissores desde a edição de 2023, com um aumento significativo na participação feminina. Além da vaga inicial proposta, foram adicionadas mais duas vagas, uma para equipes com pelo menos 2 meninas e outra para equipes com pelo menos uma menina, que não se classificariam organicamente, mas que obtiveram bom desempenho nessa nova categoria de vagas. 

Na final nacional de 2022 da Maratona de Programação, havia apenas 9 competidoras. Com a implementação das ações afirmativas, que incluíram 14 meninas, o movimento de crescimento está ganhando impulso, com um total de 24 meninas participando. Esses resultados são particularmente encorajadores, uma vez que representam um movimento de crescimento em um cenário que permaneceu estático em relação à representatividade feminina por mais de duas décadas.

\noindent \textbf{\textit{Encontros de Mulheres na Maratona:}} Em 2023 na final nacional da Maratona de programação foi realizado o primeiro encontro de mulheres na maratona em que todas as participantes foram convidadas. Os temas motivação da participação feminina, acolhimento e incentivo foram debatidos. Houve ainda, um convite, com incentivo financeiro, inclusive, à participação feminina na escola de verão 2024, evento este em que são preparados os competidores para a etapa Latino Americana da Maratona de programação. 

Em 2023 e 2024 aconteceram encontros presenciais e on-line, com a criação do grupo Meninas da Maratona. O primeiro encontro foi na final nacional em 2023, com a criação de um grupo de mensagens para troca de experiências, oportunidades e informações. O encontro na escola de verão 2024 fomentou o incentivo ao acolhimento, com palestras de mulheres inspiradoras e empresas incentivadoras. Estas ações fazem parte do conjunto de ações afirmativas construídas para o aumento da representatividade feminina nas Maratonas de Programação e na Computação.

\noindent \textbf{\textit{Educação como caminho para a inclusão:}} Além dessas iniciativas, estão sendo desenvolvidas propostas para criação de atividades formativas que visam a aumentar o desempenho de alunos e alunas de instituições superiores, preparando-os para melhorar os resultados na programação competitiva. 

Uma abordagem que enfatiza a importância da educação como o caminho principal para a inclusão e a promoção da diversidade nas competições de programação é o treinamento para a programação competitiva desenvolvido pela Universidade de Uberlândia. Equipes de todo o Brasil podem participar de competições de maneira remota, semanalmente e com temas variados, como forma de preparação para as Maratonas. Há apoio aos treinadores e grupos de discussão sobre as questões propostas em cada competição.

\section{Conclusões e trabalhos futuros} \label{sec:conclusionpage}
Esta pesquisa procurou compreender o cenário feminino na programação competitiva e os resultados das estratégias desenvolvidas para aumentar a participação feminina na Maratona de Programação SBC. Percebe-se, nesta trajetória, que a participação feminina necessita de um suporte desde a família, escola e políticas de empoderamento feminino, assim como acolhimento e propostas acadêmicas de preparação e treinamento. Só assim será possível a construção de um ambiente de equidade. 

Essas iniciativas representam um passo importante na direção da equidade de gênero na programação competitiva. Elas demonstram o compromisso em enfrentar a disparidade de gênero e garantir que todas as pessoas, independentemente do gênero, tenham igualdade de oportunidades na Computação. 

A contribuições desta pesquisa para a área de Computação e  educação em Computação é instigar e propor ações afirmativas para aumento da participação feminina brasileira nas Maratonas de Programação, pensando e repensando possibilidades no agir. Como trabalhos futuros, o desenvolvimento de programas de capacitação e mentoria, encontros de incentivo e acolhimento, assim como treinamentos de alto nível são ações serão desenvolvidas para o fomento da participação feminina.

\section{Agradecimentos}
Agradecemos ao PPGCO da UFU pelo incentivo, ao comitê da Maratona de Programação, à SBC e ao ICPC pelo fornecimento dos dados das Maratonas de programação, bem como ao apoio das ações afirmativas, em especial ao Carlinhos e Ricardo Anido. Também ao Rafael Schouery e a Lucy Mari pelo apoio às iniciativas dos encontros de Mulheres e MFP.

\bibliographystyle{sbc}
\bibliography{wei}

\end{document}